\documentstyle[11pt,aaspp4]{article}

\def\etal{et al.\ }
\def\nd{\nodata}               
\def\Deg{\hbox{${}^\circ$\llap{.}}}

\def\deg{\hbox{${}^\circ$}}
\def\min{\hbox{${}^{\prime}$}}
\def\sec{\hbox{${}^{\prime\prime}$}}
\def\kms{km s$^{-1}$}

\lefthead{Ferrarese \etal}
\righthead{A database of Pop II Secondary Distance Indicators.}

\begin{document}

\title{A Database of Cepheid Distance Moduli and TRGB, GCLF, PNLF and
SBF Data Useful for Distance Determinations}

\centerline{Accepted for publication in the {\it Astrophysical Journal Supplement Series}}

\author{Laura Ferrarese\altaffilmark{1,2}
Holland C. Ford\altaffilmark{3},  
John Huchra\altaffilmark{4},
Robert C. Kennicutt, Jr.\altaffilmark{5},  
Jeremy R. Mould\altaffilmark{6},  
Shoko Sakai\altaffilmark{7}
Wendy L. Freedman\altaffilmark{8}, 
Peter B. Stetson\altaffilmark{9}, 
Barry F. Madore\altaffilmark{10}, 
Brad K. Gibson\altaffilmark{11}, 
John A. Graham\altaffilmark{12}, 
Shaun M. Hughes\altaffilmark{13}, 
Garth D. Illingworth\altaffilmark{14},  
Daniel D. Kelson\altaffilmark{12},
Lucas Macri\altaffilmark{4}, 
Kim Sebo\altaffilmark{6}
\& N.A. Silbermann\altaffilmark{10}  
}

\altaffiltext{1}{Hubble Fellow}
\altaffiltext{2}{California Institute of Technology, Pasadena CA 91125, USA}
\altaffiltext{3}{Johns Hopkins University and Space Telescope
Science Institute, Baltimore MD 21218, USA}
\altaffiltext{4}{Harvard Smithsonian Center for Astrophysics, Cambridge MA 02138
USA}
\altaffiltext{5}{Steward Observatories, University of Arizona, Tucson AZ 85721,
USA}
\altaffiltext{6}{Research School of Astronomy \& Astrophysics, Institute of
Advanced Studies, ANU, ACT 2611, Australia}
\altaffiltext{7}{Kitt Peak National Observatory, NOAO, Tucson AZ 85726, USA}

\altaffiltext{8}{Carnegie Observatories, Pasadena CA 91101, USA}
\altaffiltext{9}{Dominion Astrophysical Observatory, Victoria, British Columbia
V8X 4M6, Canada}
\altaffiltext{10}{NASA/IPAC Extragalactic Database and California Institute of
Technology, Pasadena CA 91125, USA}
\altaffiltext{11}{CASA, University of Colorado, Boulder, CO, USA}
\altaffiltext{12}{Department of Terrestrial Magnetism, Carnegie Institution of
Washington, Washington DC 20015, USA}
\altaffiltext{13}{Royal Greenwich Observatory, Cambridge CB3 OHA, UK}
\altaffiltext{14}{Lick Observatory, University of California, Santa Cruz CA 95064
USA}

\begin{abstract}

We present a  compilation of Cepheid distance moduli and data for four
secondary distance indicators that employ stars in the old stellar
populations: the planetary nebula luminosity function (PNLF), the
globular cluster luminosity function (GCLF), the tip of the red giant
branch (TRGB), and the surface brightness fluctuation (SBF)
method. The database includes all data published as of July 15,
1999. The main strength of this compilation resides in all data being
on a {\it consistent and homogeneous} system: all Cepheid distances
are derived using the same calibration of the period-luminosity
relation, the treatment of errors is consistent for all indicators,
measurements which are not considered reliable are excluded. As such,
the database is ideal for inter-comparing any of the distance
indicators considered, or for deriving a Cepheid calibration to any
secondary distance indicator, such as the Tully-Fisher relation,
the Type Ia Supernovae, or the fundamental plane for elliptical
galaxies. This task has already been undertaken by Ferrarese \etal
(2000a), Sakai \etal (2000), Kelson \etal (2000) and Gibson \etal
(2000). Specifically, the database includes: 1) Cepheid distances,
extinctions and metallicities; 2) reddened apparent $\lambda$5007 \AA~
magnitudes of the PNLF cutoff; 3) reddened, apparent magnitudes and
colors of the turnover of the  GCLF (both in the $V-$ and $B-$bands);
4) reddened, apparent magnitudes of the TRGB (in the $I-$band) and
$V-I$ colors at and 0.5 magnitudes fainter than the TRGB; 5) reddened,
apparent surface brightness fluctuation magnitudes measured in
Kron-Cousin $I$, $K'$, $K_{short}$, and using the F814W filter with
the $HST$/WFPC2. In addition, for every galaxy in the  database  we
give reddening estimates from DIRBE/IRAS as well as HI maps, J2000
coordinates, Hubble and T-type morphological classification,
apparent total magnitude in $B$, and systemic velocity. 

\end{abstract}


\section{Introduction}

The Hubble Space Telescope ($HST$) Key Project on  the Extragalactic
Distance Scale (Kennicutt, Freedman \& Mould 1995, Ferrarese \etal
1999) as well as other $HST$ programs (Saha \etal 1995, 1996a, 1996b,
1997, 1999, Tanvir \etal 1995),  have provided Cepheid distances to 24
galaxies in 11 clusters as far as Virgo and Fornax. Added to the
existing ground-based Cepheid distances to 10 Local Group galaxies,
they constitute a solid background against which to calibrate
secondary distance indicators.  To do so, the first step is to collect
the available data, which are often inhomogeneous in terms of adopted
zero point, extinction correction, and error analysis, and set them on
a common scale. In this paper we accomplish this task for the
published Cepheid distances as well as for four  distance indicators
that employ stars in the old stellar populations (Pop II):  the tip of
the red giant branch (TRGB), the planetary nebula luminosity function
(PNLF), the globular clusters luminosity function (GCLF) and the
surface brightness fluctuation  method (SBF).  All data available in
the literature as of July 15, 1999 are reviewed.  While this paper deals
exclusively with the issue of compiling a homogeneous database, a
second paper (Ferrarese \etal 2000a) uses the database to provide a
Cepheid based calibration for the above-mentioned distance
indicators, and employs SBF distances to step out into the Hubble flow
and derive a value for the Hubble constant. 

Extensive reviews of the TRGB, PNLF, GCLF and SBF, and their
application as distance indicators have been published by Madore \etal
(1996), Sakai (1999), Jacoby \etal (1992), Jacoby (1996), Jacoby
(1998), Whitmore (1996), Tonry \etal (1996) and Blakeslee,
Ajhar \& Tonry (1998). All methods are also reviewed critically in
Ferrarese \etal (2000a), where relative strengths and weaknesses are
discussed. This paper is organized as follows. The next sections will
describe the database (\S 2), outline the criteria used in
homogenizing the published data and discuss cases that needed special
attention  (\S 3 to \S 7).  Relative distances to galaxies in groups
and clusters allow the study of the clusters' geometry. Furthermore,
a Cepheid calibration or a comparative  study of any secondary
distance indicators can be  accomplished either by using a direct
comparison or  group mean distances. Therefore we discuss the
group/cluster membership for every galaxy in the database in Appendix
A.

\section {Description of the Database}

The task of compiling an internally consistent database is achieved by
1) assuring that Cepheid distance moduli share the same calibration
and a common procedure in fitting the period-luminosity (PL)
relation; 2) publishing reddened, characteristic magnitudes rather
than distance moduli for the TRGB, PNLF, GCLF and SBF methods: listed
in the database are the $I-$band magnitudes of the TRGB, $I_{TRGB}$,
the cutoff magnitude of the PNLF, $m^*$ (measured in the
[OIII]$\lambda$5007 emission), the turnover magnitude of the GCLF,
$m_T$, and the fluctuation magnitude for SBF, $\overline{m}$, all of
which are not corrected for Galactic reddening; 3) considering a
consistent  error analysis for each indicator.

We review all Cepheid distance moduli, and all TRGB, PNLF, GCLF, and
SBF data published in refereed journals as of July 15, 1999. In some
cases, slight changes to the published data were necessary to conform
to the criteria given above, these cases are discussed in \S 3 through
\S 7.  The $I-$band SBF (hereafter $I$-SBF) dataset of Ajhar \etal
(1999), is finalized but not yet published at the time this paper
is written, and it has not been duplicated here in its
entirety. Rather, we include in this paper only $I$-SBF fluctuation
magnitudes to galaxies with at least one other Cepheid, TRGB, GCLF or
PNLF measurement. Reddening corrected and calibrated PNLF, GCLF, SBF
and TRGB distance moduli for all galaxies considered in this paper can
be found in Ferrarese \etal (2000a).
  
The database consists of three tables. Table 1 gives general
information for all galaxies with at least one Cepheid, TRGB,
PNLF, GCLF, or SBF measurement (with the above-mentioned  exception of the
$I$-SBF database of Ajhar \etal). The table is organized as follows: Column 1: galaxy name;
columns 2 and 3: right ascension and declination at J2000; columns 4
and 5: Hubble Type and  T-type respectively; column 6: total
apparent  $B$ magnitude, $B_T$; column 7: heliocentric systemic
velocity (in \kms). All entries are from the RC3 
(de Vaucouleurs \etal 1991) except for the
heliocentric velocity, which is taken from various sources as listed by
the NASA Extragalactic Database (NED). The galaxies are organized in
order of increasing right ascension.

For the same galaxies listed in Table 1, Table 2 includes reddening
and metallicity indicators appropriate to the stellar population
employed by each method. Three reddening estimates are given: the
total (Galactic plus internal) reddening determined from observations
of Cepheids (when available) in column 2; the Galactic reddening based
on the DIRBE/IRAS maps of Schlegel \etal  (1998) in column 3; the
Galactic reddening based on the HI maps of Burstein \& Heiles (1984)
in column 4. From the data listed in the table, reddening correction
can be performed easily once an extinction curve is adopted: for
example, for $R_V=3.3$, the extinction curves of Cardelli, Clayton \&
Mathis (1989)  give the following ratios for the absorption:
$A(B):A(5007):A(V):A(I):A(F814W):A(K_{s}):A(K')$ =
1.288:1.120:1:0.600:0.596:0.120:0.125.  The metallicity indicators
listed in Table 2 are as follows: the metallicity for the old stellar
population in elliptical galaxies and bulges is measured by the Mg$_2$
indexes (column 5) and by the $V-I$ colors determined in parallel to
the SBF measurements (column 10); globular cluster metallicities are
represented by the globular cluster color determined at a turnover
magnitude of the GCLF (column 6); the Cepheid metallicity is given in
column 7 as the [O/H] index for the HII regions in the immediate
vicinity of the Cepheids;  finally the metallicity of the halo can be
estimated from the $V-I$ colors determined at the TRGB (column 8), or
more precisely at an absolute $I$ magnitude of $\sim -3.5$, or 0.5
magnitudes fainter than the TRGB (column 9). References are given in
the notes to the table.

Cepheid distance moduli and magnitudes for the secondary distance
indicators are listed in Table 3. Here, the galaxies are organized
according to group membership, as defined in Appendix A. The  columns
are as follows. (1): galaxy name; (2), (3), (4) and (5): the number of
Cepheids used in determining the PL relation, the photometric bands in
which the Cepheid observations were carried out, the Cepheid distance
modulus and references respectively; (6) and (7) the apparent $I$
magnitude of the TRGB not corrected for Galactic extinction, and
references; (8), (9) and (10): the number of PNe used in fitting the
PNLF,  the apparent PNLF cutoff magnitude, uncorrected for extinction,
and references; (11) and (12): the apparent  turnover magnitude of the
GCLF, uncorrected for extinction and determined in the photometric
band listed next to each measurement, and references; (13) and (14):
the SBF fluctuation magnitude, uncorrected for extinction, and
determined in the photometric band listed next to each measurement,
and references. When quoting data from this paper, the references
given in the table should be cited.

In Tables 2 and 3, data followed by a colon are reported for
completeness, but not considered reliable for the purpose of
calibrating the distance indicators, for the reasons outlined in the
following sections.

\section{Cepheid Distances}

Table 3 lists published Cepheid distances. When multiple measurements
to the same galaxy exist,  we only list the most comprehensive (and
generally most recent) study.

To insure homogeneity within the sample, it is necessary that:  a) all
Cepheid distances rely on the same absolute calibration, i.e. adopt a
common zero point  and slope for the period-luminosity (PL)
relation; b) possible incompleteness biases affecting the PL relation are
taken into account; c) extinction corrections are performed according
to a common reddening law and the same value of the ratio of
total to selective absorption $R_V$; and d) a common and complete list
of uncertainties is factored in when calculating the error in the
quoted distances.

The PL calibration adopted in this paper is based on a sample of 32
LMC Cepheids with  $BVI$ photoelectric photometry and periods in the
range from 1.5 to 63 days (Madore \& Freedman 1991). The absolute
calibration assumes a true distance modulus and average line-of-sight
reddening  to the LMC of $18.50 \pm 0.13$ mag (Mould \etal 2000) and $E(B-V) =
0.10$ mag respectively,  a ratio of total to selective absorption
$R_V=A(V)/E(B-V)=3.3$, and a reddening law following Cardelli, Clayton
and Mathis (1989). This calibration is shared by  all but two of the
published Cepheid distances.  The  exceptions are the Small Magellanic
Cloud (Welch \etal 1987) and  NGC 3109 (Musella \etal 1997), and are
discussed in \S 3.2 and \S 3.3.

In their discussion of the distance modulus to the Virgo cluster
galaxy M100, Ferrarese \etal (1996) performed a detailed analysis of
the bias arising from magnitude incompleteness at short periods. It
was concluded that the Cepheid sample in M100 is substantially
incomplete below 20 days.  Because of the finite width of the Cepheid
PL relation, only the brighter  Cepheids are recovered at the faint
end of the PL relation, leading to an underestimate of the distances
by 0.05-0.10 magnitudes if the bias is not corrected.  These
conclusions are further supported by artificial star tests to be
discussed in Ferrarese \etal (2000b). At the distance modulus
of the Virgo and Fornax clusters, these  tests show that the $V$ PL
relation is  $\sim$50\% incomplete below 20$-$25 days. Because of
these results, we found it necessary to revisit the $HST$ Cepheid
distances published as part of the Key Project. For each galaxy we
calculated distance moduli for samples with different lower cutoff in
period.  In a few cases (NGC 1365, NGC 1326A, NGC 1425, NGC 4725 and
NGC 4535) we found that the distance modulus increases when the lower
period cutoff is moved from $\sim10$ to $\sim 20 - 25$ days, the exact
period depending on the distance and level of crowding in the
galaxy. The period cutoff above which the distance moduli stabilize is
then applied to the data. This produces an increase in the
derived distance modulus of  $\sim 0.1$ mag at most.  Changes made to
published distance moduli are detailed in Table 4. Notice that a
lower period cutoff was not applied to the LMC Cepheid sample
that calibrates the PL relation, however, because the slope of the PL
relation fitted to each distant galaxy is kept fixed to the LMC value,
this does not introduce a bias in the derived distance moduli
(e.g., Graham et al. 1999).

To allow for meaningful weighting, all published error estimates have
been revised slightly to include only random contributions.  A typical
example of the error budget adopted is presented in Ferrarese \etal
(2000a). Random sources of error include scatter in the fit of the
PL relation, and the error on the assumed value of the ratio of total
to selective absorption, $R_V$. Systematic sources of error include
the uncertainty in the distance to the LMC and the error in the zero
point of the LMC PL relation.  For the Cepheid distances based on
$HST$ photometry, we have treated the photometric error  as systematic,
rather than random; therefore  the sum, in quadrature, of all of the
above systematic errors, amounts to 0.16 mag, and has not been listed
explicitly in Table 3.  For the Cepheid galaxies based on
ground-based photometry, the only  sources of systematic error are
the uncertainty in the LMC distance and the LMC PL relation zero
point, 0.13 magnitudes; the photometric errors have been treated as
random. It follows that if a sample of Cepheid is used  including {\it
both} ground-based and $HST$ observations, a 0.14 mag uncertainty
(due to photometric zero points errors) must be added in quadrature  to the error
reported in Table 3 for the $HST$ galaxies, and the systematic error
for the entire sample would be reduced to 0.13 mag.

Distance uncertainties arising from  a possible metallicity dependence
of the Cepheid PL relation (e.g. Kennicutt \etal 1998) are not
accounted for at this stage.  Metallicity effects on the calibration
of secondary distance indicators are considered explicitly in
Ferrarese \etal (2000a), Sakai \etal (2000), Kelson \etal (2000),
and Gibson \etal (2000).  

The reddening derived from multi-wavelength observations of Cepheids is
listed in column 2 of Table 2 as $E(B-V)=A_V/3.3=0.742 \times
E(V-I)$. Errors are associated with the reddening only when given by
the authors. If Cepheids are observed in only one photometric band, a
N/A appears in column 2. Note that $E(B-V)$ includes both foreground
(Galactic) and internal extinction, and should therefore be larger
than the  Schlegel \etal (1998) and Burstein \& Heiles (1984)
$E(B-V)$ values given in columns 3 and 4 of Table 2.

\subsection{Rejected Cepheid Distances}

Sandage \& Carlson (1985) published periods and light curves for 15
Cepheids in the WLM galaxy. Their photometry was subsequently 
questioned by Ferraro \etal (1989); furthermore, all of the WLM
Cepheids have periods less than 10 days, and might therefore be
contaminated by overtone pulsators. Therefore we do not include the
WLM distance in the database.

Cepheid distances based on observations in a single photometric band
are listed in the database as upper limits, because of the
impossibility of deriving a mean internal reddening (Freedman \&
Madore 1988). These galaxies include Holmberg II (=UGC 4305), NGC 2403, 
GR8, NGC 2366 and NGC 4571.

\subsection{Cepheid Distance to the Small Magellanic Cloud}

Preceding the Madore \& Freedman PL calibration, Welch \etal (1987)
derived $(m-M)_{SMC} = 18.93 \pm 0.05$ mag and negligible extinction
for the Small Magellanic Cloud,  from $HJK$ photometry of  91
Cepheids.  Using the same data,  and adopting zero points and slopes
derived for the IR PL relations by Madore and Freedman
(1991), we obtain apparent distance moduli
$(m-M)_J=18.997 \pm 0.024$, $(m-M)_H=19.013 \pm 0.022$, and
$(m-M)_K=18.989 \pm 0.022$, where the errors reflect only the rms
scatter around the mean PL relation. Using the Cardelli, Clayton and
Mathis (1989) reddening law with $R_V = 3.3$, the true distance
modulus and reddening to the SMC are $(m-M) = 18.99 \pm 0.05$ and
$E(B-V)=0.01 \pm 0.05$. The error on the distance modulus accounts for
a 3\% error in the photometry (Welch \etal 1987), the scatter in
the PL relations, and the (negligible) error in the assumed value
of $R_V$.

\subsection{Cepheid Distance to NGC 3109}

The most recent and complete data set for Cepheid variables in NGC
3109 were presented by Musella, Piotto \& Capaccioli (1997). Sixteen
of their 36 Cepheids (with periods in the range 2.8 to 20 days) have
$BVRI$ measurements.  A distance modulus $(m-M)=25.67 \pm 0.16$, and
reddening $E(B-V)=-0.01 \pm 0.04$ are derived using a calibrating PL
relation based on  Sandage (1988) and Madore \& Freedman (1991), and
distance modulus and reddening for the LMC of $18.50$ mag and
$E(B-V)=0.08$ mag respectively.

We derived a distance to NGC 3109 by applying the standard procedure
described in \S3. We therefore selected only Cepheids with period
larger than 10 days and observed in all available passbands,
$BVRI$. Unfortunately, this  reduces the sample to only four Cepheids,
with the consequence that the distance cannot be constrained too
well. The exclusion of short period Cepheids is necessary in order to
avoid overtone pulsators and the possibility of magnitude
incompleteness: from Figures 4$-$7 of Musella \etal, Cepheids with
periods shorter than 10 days clearly define a brighter PL relation
than Cepheids with longer period. Our procedure yields apparent
distance moduli $(m-M)_B=25.97 \pm 0.18$, $(m-M)_V=25.91 \pm 0.14$,
$(m-M)_R=25.71 \pm 0.11$, and $(m-M)_I=25.60 \pm 0.09$. From this we
obtain a formal true distance modulus $(m-M)=25.26 \pm 0.22$ mag, and
a reddening $E(B-V)=0.18 \pm 0.08$ mag. This result is obtained by
using weights proportional to the inverse square of the errors on the
individual apparent distance moduli, not using weights would increase
the distance modulus by 0.05 mag. In the particular case of NGC 3109,
for which most of the observed Cepheids are at short period, a better
estimate of the reddening could be obtained by calculating the
reddening  for all of the Cepheids on an individual basis, rather than
for the combined long period subset. This procedure is however not
standard for the galaxies observed by the Key Project and would
introduce inhomogeneities. The adoption of a different PL calibration,
and the exclusion of short period Cepheids contribute in equal amount
to the difference between our distance and Musella \etal

\section{Tip of the Red Giant Branch Distances}

For distance determinations, the magnitude of the TRGB is always
measured in the $I-$band, since this is where metallicity effects are
minimized (Lee, Freedman \& Madore 1993). Metallicity corrections can
be estimated from the $V-I$ color at the TRGB and 0.5 mag fainter
(corresponding roughly to an absolute magnitude $M_I=-3.5$). These, uncorrected
for foreground extinction, are listed in columns 8 and 9 of Table 2,
with the exception of NGC 3379 (Sakai \etal 1997), and IC 3388 (Harris
\etal 1998), for which only $I$ data were obtained.

The TRGB $I_{TRGB}$ magnitude (uncorrected for extinction) and
associated error are listed in column 6 of Table 3.  Most authors 
include the internal uncertainty of the fit to the TRGB in the error
quoted for  $I_{TRGB}$, and incorporate photometric errors in the
final distance modulus. Therefore, to obtain a consistent dataset,
photometric errors (as quoted by the authors) and uncertainties in the
TRGB fit were added in quadrature to produce the $I_{TRGB}$ error
listed in Table 3.

\section{Planetary Nebulae Distances}

The $\lambda$5007 \AA~cutoff magnitude of the PNLF, $m^*$, uncorrected
for foreground extinction, is listed in column 9 of Table 3.  Because
PNLF distance moduli are calculated by fitting the luminosity function
with a standard template (e.g. Ciardullo \etal 1989), only the final
distance moduli are published. From these we derived $m^*$  {\it a
posteriori} by subtracting the zero point (and the extinction
correction if applied) adopted by the authors.  The PNLF distances to
the SMC, NGC 3109, and NGC 5253, listed in Table 3, are not well
constrained. The PNe sample in NGC 3109 (Richer  \& McCall 1992)
includes only seven objects, and an upper limit to the distance is
derived from the brightest of the PNe observed. Jacoby, Walker \&
Ciardullo (1990) advise against the use of the PNLF distance to the
SMC due to the small number of PNe defining the luminosity function.
Finally, the small number of PNe detected in NGC5253, the presence of
strong internal dust extinction, and the galaxy's very low metal
abundance all conjure to produce a very ill constrained PNLF magnitude
cutoff, unsuitable for distance determinations (Phillips \etal 1992).

Uncertainties in the values of $m^*$ are summarized, for example, in
Jacoby,  Ciardullo \& Ford (1990). They include a contribution
associated with the fitting procedure (of the order of 0.10 mag),
photometric zero points ($\sim 0.05$ mag),  the filter response
calibration ($\sim 0.04$ mag), and the uncertain definition of the
empirical PNLF ($\sim0.05$ mag). Errors in the reddening estimate,
which are sometimes included, have been removed (in quadrature) from
the present analysis, as we only deal with uncorrected magnitudes.

\section{Globular Clusters Luminosity Function Distances}

The task of building a homogeneous data sample from the available GCLF
data is far from simple.  Most of the published data are in the $V-$band, 
or in the $C$, $T_1$ Washington system which can readily be
transformed to give $V$ magnitudes (Geisler 1996).  However,  there
are several cases in which only the $B-$band luminosity function is
presented. 

The turnover magnitude cannot be properly estimated unless the
luminosity function extends beyond it (Ferrarese \etal 2000a); 
galaxies with a GCLF not sampling the turnover
are therefore not included in our database. Unfortunately, this
excluded NGC 720 (Kissler-Patig \etal 1996), NGC 4636 (Kissler \etal
1994), the Fornax sample of Kohle \etal (1996), the Virgo 
cluster sample
of Ajhar \etal (1994), NGC 1400 and NGC 1407 in the Eridanus cluster
(Perrett \etal 1997), and NGC 4881 in Coma (Baum \etal 1995).  The
GCLF for NGC 1399 published by Bridges, Hanes \& Harris (1991)
presented us with a difficult choice. The authors fit the $V$ GCLF
using fixed $\sigma$ in the range $1.00 \leq \sigma \leq 1.75$ and
find that $\chi^2$ is minimized for $\sigma \sim 1.5$, giving a
turnover at $V=24.10$, only 0.3 magnitudes fainter than their
magnitude limit.  In $B$, however, the $\chi^2$ of the fit keeps
decreasing while sigma is increased, up to the last fitted value of
$\sigma=1.75$, which gives a turnover half a magnitude fainter  than
the magnitude cutoff of the data. There are not enough data in the
literature to determine whether the GCLF dispersion in $B$ is
intrinsically larger than in $V$, and therefore to decide whether the
Bridges \etal results are simply due to the $B$ data being more
incomplete than (not as deep as) the $V$ data, or to the fact that the
turnover magnitude is not sampled well enough to determine an accurate
distance. For these reasons, we   conservatively excluded the  Bridges
\etal  data from  our sample. Finally, the Elson \& Santiago (1996)
study of globular clusters in M87 was not considered because the
authors warn that the difficulty in estimating the contamination from
background galaxies makes  the determination of the turnover magnitude
very uncertain and not suitable for distance determinations. 

Four cases  required further attention.  The NGC 3031 data of
Perelmuter \& Racine (1995) do extend beyond the turnover, but the
procedure  used to analyze the data is unclear. We could not determine
whether the authors solved for $\sigma$ or not. We  performed a  three
parameter Gaussian fit to the data, and found $V=20.86\pm0.46$ and
$\sigma=1.46\pm0.25$. This is significantly different from the
published value of $V=20.3\pm0.3$.  The source of this difference is
unclear, and the data have large errorbars; consequently, we decided to
exclude the GCLF distance to NGC 3031 from our sample.

In the case of NGC 3377,  Harris (1990) found it necessary to combine
the GCLF with that of NGC 3379, assumed to be at the same distance.
This produces a sufficiently well defined GCLF for a three parameter
fit.  We found that a three parameter fit to the NGC 3377 GCLF alone
is possible even if rather shaky. Our fit gives $B=23.13 \pm 0.32$ and
$\sigma=0.88 \pm 0.22$ for NGC 3377, in reasonably good agreement with
the Harris values of $B=23.35 \pm 0.40$ for the combined NGC 3377 +
NGC 3379 GCLF. However, we follow Harris' conclusion that a
fit to NGC 3377 alone is not reliable, and we do not include this
measurement in our table.

We also performed our own fit to the Fleming \etal (1995) data for
NGC 4494. In spite of the fact that the GCLF does sample the turnover,
the authors fixed the value of $\sigma$ in fitting the data.  We found
$V=23.32\pm0.35$ and $\sigma=1.31\pm0.18$, and adopt it in our Table
3.  This compares to the published $V=23.6\pm0.4$, which assumes a
$\sigma$ of 1.4.

Finally, for IC 4051 in Coma, we quote in Table 3 the best fit to the
turnover reported by Baum \etal (1997) using a Gaussian fit. The
authors do not give an error for this measurement; consequently, we
adopted half the difference between the turnover magnitudes obtained
by Baum \etal  by fitting a Gaussian and an asymmetric hyperbolic
function as an estimate of the uncertainty. It is not entirely clear
whether the dispersion of the Gaussian was allowed to vary in the fit,
and we cannot reproduce the result because the data are not tabulated. 

In keeping with our precepts, the GC data reported in Tables 2 and 3
are not corrected for either internal or Galactic reddening. The
turnover magnitudes are listed in column 11 of Table 3, and the GC
color at the turnover, when available, can be found in column 6 of
Table 2.  Photometric errors are always small compared to the formal
uncertainty in the Gaussian fits to the GCLF, and have not therefore
been taken into account: the quoted errors for the turnover magnitudes
in Table 3 include only the formal uncertainty in the fit. There are,
of course, exceptions. $HST$ data often comprise large samples of GCs
extending well beyond the turnover of the luminosity function, and the
fits are very well constrained. Therefore, we added (in quadrature) a
0.06 magnitude uncertainty to the fitting error quoted for NGC 1399,
NGC 1404, NGC 3113, NGC 4486, NGC 5846, NGC 4494 and NGC 4278
(Grillmair \etal 1999, Kundu \& Whitmore 1998, Forbes 1996a, Forbes \etal
1997, Kundu \etal 1998). This accounts for a 0.05 mag error in
the photometric zero point, plus a 0.03 magnitude error in the
photometric aperture correction (Holtzman \etal 1995).

A final note of caution is necessary. Measuring the peak of GCLFs is
very tricky: it requires precise estimates of 1) the incompleteness
of the data, 2) the errors as a function of magnitude (that will
`scatter' points preferentially to brighter magnitudes), and 3) the
contribution of the background galaxy population.   In the worse case,
all these effects can induce a 0.5 magnitude error if not taken into
account. We did not attempt to quantify or check any of these effects;
instead we rely on the authors' analysis.

\section{Surface Brightness Fluctuation Distances}

The surface brightness fluctuation method has been employed in ground
based surveys using the Kron-Cousins $I-$band (Ajhar \etal  1999)
and the near-infrared $K'$ ($\lambda_c = 2.10 \mu$m) and $K_{s}$
($\lambda_c = 2.16 \mu$m) filters (Jensen \etal 1996, Pahre \& Mould
1994). Several galaxies have been observed  with the F814W filter
using the $HST$/WFPC2 (Ajhar \etal 1997, Lauer \etal 1998).  Because
of the dependence of the fluctuation magnitude on the stellar
population, SBF observations in different passbands are listed
separately in Table 3 (column 13).  All fluctuation magnitudes include
a $k-$correction term, but are not extinction corrected. Errors on the
quoted magnitudes and colors include photometric zero points errors,
and the estimated error on the fluctuation magnitudes.

The calibration of the SBF method includes a strong metallicity
dependence (Tonry \etal 1997), which can be quantified through the
mean $V-I$ color of the region in which the fluctuations are
determined. This is listed in column 10 of Table 2, also uncorrected
for extinction, but including a $k-$correction term.

\section{Summary}

We have presented a homogeneous compilation of Cepheid, TRGB, PNLF,
GCLF and SBF data that can be used in determining distances to
external galaxies.  The database represents the most complete and
coherent collection of measurements for both the Cepheids and the four
Pop II distance indicators as of July 15, 1999.  All data, with the
exception of the Cepheid distance moduli, are uncorrected for Galactic
extinction, estimates of which are given based both on DIRBE/IRAS maps
(Schlegel \etal 1998) and HI maps (Burstein \& Heiles 1984). Reddening corrected
distance moduli are given for all of the galaxies in the database in
Ferrarese \etal (2000a), using a calibration based on Cepheids. Cepheid
distances presented in this paper  have also been  used for the
calibration of the Tully-Fisher relation (Sakai \etal 2000), Type
Ia supernovae (Gibson \etal 2000), and the fundamental plane of
elliptical galaxies (Kelson \etal 2000)
	
We wish to thank John Tonry, John Blakeslee, Ed Ajhar and Alan Dresser
for kindly giving us access to the SBF database prior to publication.
LF acknowledges
support by NASA through Hubble Fellowship grant HF-01081.01-96A
awarded by the Space Telescope Science Institute, which is operated by
the Association of Universities for Research in Astronomy, Inc., for
NASA under contract NAS 5-26555.  Support
for this work   was also provided by NASA  through grant GO-2227-87A
from STScI.
This research has made use of the NASA/IPAC Extragalactic Database (NED),
version 2.5 (May 27, 1998). NED is operated by the Jet Propulsion Laboratory, 
California Institute of Technology, under contract with the National 
Aeronautics and Space Administration.

\appendix

\section{Identification of Groups and Cluster}

A complete compilation of galaxies belonging to different groups and
clusters is beyond the scope of this paper. Our goal is to establish
physical associations between galaxies with a measured Cepheid
distance and at least one between PNLF, GCLF, TRGB, and SBF distance
estimates.  Group membership for all of the galaxies of interest has
been analyzed using the CfA redshift catalogue (zcat, version July
1998, Huchra \& Mader 1998). Criteria for group membership vary from
group to group: a detailed description of the methodology adopted is
given below for each  case.  When possible, our analysis has been
supplemented with published studies of galaxy groups, as listed
below. For each identified group, galaxies are divided into three
classes: certain members (galaxies for which group membership is
beyond any doubt, based on position, redshift and, when available,
previous population studies of the group), probable members (galaxies
that are most likely true members, but are slightly too far, in either
space or redshift, from the certain members of the group to pass the
`beyond any reasonable doubt' condition, according to criteria
established below for each group), and possible members (galaxies that
are probably not true members, but for which the association with the
group cannot be completely excluded). Finding charts for all galaxies
in each group are given in Figures 1a and 1b, and cluster properties
are given in Table 5.

\noindent {\bf The M81 Group.}  An inspection of zcat shows that
the galaxies in the M81 group are divided in two separate sub-clumps: the
first extends in a 4\deg$\times$2\deg~region centered on M81, while
the second is located about 13\deg~to the east. The core of the group,
defined by M81, NGC 3034 and NGC 3077, is very compact: the three
galaxies are found within 1\deg~and 100 \kms~of each other.  IC 2574
and NGC 2976 are located 2\deg~to 3\deg~from M81, and have comparable
velocity.  The second sub-clump, 13\deg~to the east of M81, is
defined by NGC 2366, NGC 2403, and 08141+7052. In view of the fact
that all the galaxies in the two sub-clumps share the same systemic
velocity within 400 \kms, we classify NGC 2366 and NGC 2403 as
probable members of the M81 group, in spite of the large separation
from the galaxies defining the group core. NGC 2787 is located 3\Deg3
west of M81.  The systemic velocity of ~700 \kms~places it at the high
boundary of the velocity distribution for the M81 group, and therefore
we classify this galaxy as a possible member.  In the Nearby Galaxies
Catalog (Tully 1988, hereafter T88), M81, NGC 3077, NGC 2403 and NGC
2366 are all identified as belonging to the Coma-Sculptor Cloud
(cloud number 14), group $-$10, while NGC 2787 is placed in the Ursa
Major Cloud (cloud number 12), group $-$0.  Of the galaxies discussed,
only M81 and NGC 3077 are part of the galaxy sample used by Nolthenius
1993 (hereafter N93) for his group compilation. Both galaxies are
listed as members of the M81 group (group number 40).  The dwarf
ellipticals BK5N and F8DI and their membership to the M81 group are
discussed by Caldwell \etal (1998) and Karachentseva (1985).

\noindent {\bf The NGC 5128 Group.}  The group is quite sparse and
extended. NGC 5102 and NGC 5253 are found 7\deg~and 14\deg~north
respectively of NGC 5128, the brightest galaxy in the group. Both NGC
5102 and NGC 5128 are isolated, while NGC 5253 is located in a
sub-clump of 3 galaxies (one of which is NGC 5236, the brightest spiral in the area), all within 2\deg$\times$2\deg~ and 100
\kms~in redshift. Because NGC 5128, NGC 5102 and NGC 5253 and NGC 5236 have very
similar systemic velocity (the total range spanned is 150 \kms~at ~500
\kms~redshift), and because of the lack of obvious sub-clumping,
physical association has been assumed. All three galaxies listed in the NGC
5128 group are classified as part of the Coma-Sculptor Cloud (cloud
number 14), group $-$15 by T88.

\noindent {\bf The M101 Group.}  M101 (NGC 5457) is the brightest
member of the group, which is very loose (as for the NGC 5128 group,
it is not possible to identify a well defined core: only two galaxies,
NGC 5477 and 13529+5409 are found within a 2\deg~radius and 1000 \kms~ in
systemic velocity). NGC 5195 is found in a small sub-clump of
three galaxies about 8\deg~south-east of M101, while NGC 5866 is
located in a sub-clump of four galaxies 10\deg~in the opposite
direction.  Because of the good agreement in systemic velocities, NGC
5195 and NGC 5866 have have been classified as possible members of the
M101 group, in spite of the large spatial separation. However, note
that both N93 and T88 place the three galaxies in different clouds or
groups: N93 classifies NGC 5866 in group No. 147, and NGC 5195 in
the M51 group (group number 115, NGC 5102 is not part of the catalog);
T88 places NGC 5457 and NGC 5195 in the Coma-Sculptor Cloud, but the
first in group $-$9, and the second in group $-$5, while NGC 5866 is
located in the Draco Cloud (cloud number 44), group $-$1.

\noindent {\bf The NGC 1023 Group.}  The group is very well
defined. All of the galaxies classified as certain members are found
within 6\deg~of NGC 1023 and have very similar systemic velocity. NGC
855, for which an SBF distance exists, is located 12\deg~away from NGC
1023, and 6\deg~from NGC 925, which is itself at the outer boundary of
the core region. The systemic velocity of NGC 855 is only 50
\kms~lower than for NGC 1023. Because no other obvious member of the
group is located outside a 6\deg~radius from NGC 1023, we classify NGC
855 as a probable, rather than certain member. Unfortunately, NGC 855
is not included in either the N93 or the T88 catalogs. T88 classifies
all of our certain members of the NGC 1023 group as belonging to the
Triangulum spur (cloud number 17), group $-$1. NGC 891 is further
placed in the first level association +1.

\noindent {\bf The NGC 3184 Group.}  The core of the group is defined
by NGC 3184, NGC 3198 and NGC 3319. These are all located within
3\deg~of each other, and show very good agreement in their systemic
velocities. NGC 3413 is located 13\deg~south-west of the
core. However, the galaxy has the same systemic velocity as NGC 3198,
and one additional galaxy with very similar velocity, NGC 3432, is
located 7\deg~SW of the group core, bridging NGC 3413 to the
center. For these reasons, and in spite of the large spatial
separation, we list NGC 3413 as a probable member of the
group. Unfortunately, NGC 3413 is not part of the T88 or N93 catalogs.
T88 places both NGC 3198 and NGC 3319 in the Leo Spur (cloud number
15), group +7.

\noindent {\bf The Leo I Group.}  All of the galaxies classified as
certain members are within 2\deg$\times$2\deg~and 230 \kms. About
8\deg~north-east, and at the same systemic velocity as the Leo I
group (as defined above) is the very tight quartet formed by NGC 3599,
NGC 3605, NGC 3607 and NGC 3608.  The SBF distances to these galaxies
place the group over one magnitude farther than Leo I, excluding
physical associations between the two groups. About 6\deg~east we find
a tight quartet formed by the late type spirals NGC 3623, NGC 3627,
NGC 3628 and NGC 3593 at the same systemic velocity as, and possibly
associated with, the Leo I group. All of the galaxies listed as certain
members of the Leo I group are also classified as such in the N93
(group number 50) and T88 (Leo Spur, group +7) catalogs.

\noindent {\bf The NGC 7331 Group.}  NGC 7331 and NGC 7457 are about
7\deg~ from each other, and have the same heliocentric velocity. T88
assigns the two galaxies to the same cloud but different groups. We
classify their physical association as probable.

\noindent {\bf The Coma I and Coma II Clouds.}  This is a very complex
region, and group association cannot be established with confidence
for any of the galaxies. The highest concentration of galaxies appears
to be at the location of NGC 4278 and NGC 4283, which have systemic
velocities of 643 \kms~and 1076 \kms~respectively. The other six
galaxies within 1\deg~of the pair have systemic velocities in between
these two extremes. NGC 4414 is standing alone 2\deg~to the
north-east, with a systemic velocity of 720 \kms. Because of its
proximity to the core, the lack of association with any other obvious
group, and the agreement in systemic velocities, we consider it
probably associated with the Coma I cloud. The same is true for NGC
4251 (systemic velocity 1067 \kms), which is located 2\deg~to the
south-west. NGC 4627 is located almost 4\deg~north-east of NGC
4414, in a group of three galaxies with systemic velocity around 700
\kms. Because of the agreement in systemic velocity we consider NGC
4627 as a likely member of the Coma I cloud. Our analysis largely
agrees with N93, who list NGC 4251, NGC 4283, NGC 4414 and NGC 4627 as
members of the Coma I cloud (group number 81), but NGC 4278 as part of
a sub-clump of five galaxies (group number 100). NGC 4150 is found
within 2\deg~of the core, but because of its low systemic velocity
(226 \kms), we classify it as a probable, rather than a certain member
of the cluster.  A small sparse group of five galaxies (NGC 4565, NGC
4494, NGC 4562, NGC 4747 and NGC 4725) lies about 6\deg~to the
south-east of the core of the Coma I cloud. All of the galaxies in
this group, which we identify as the Coma II cloud, have systemic
velocities in the narrow range 1190 \kms~to 1395 \kms, outside the
range spanned by the galaxies identified as certain members of the
Coma I cloud (600 to 1000 \kms).  The group extends for about 4\deg~in
the east-west direction.  All of the galaxies which are here
classified as part of the Coma I or II clouds are placed in the
Coma-Sculptor Cloud, group $-1$, by T88, with the exception of NGC
4627 (which is not included in the catalog) and NGC 4725, which is
placed in group $-$2, first level association +1.

\noindent {\bf The Virgo Cluster.}   The structural complexity of the
Virgo cluster has been recognized and studied for more than thirty
years (e.g. de Vaucouleurs \& de Vaucouleurs, 1973). We limit our
analysis to the regions where galaxies with Cepheid distances are
found. For these regions, Table 6 summarizes the spatial and
kinematical structure identified by different authors within the
cluster. In spite of the complication introduced by the fact that a
succession of authors have used different criteria and different
nomenclature to describe the structure within the  cluster, all agree
on identifying two  prominent sub-structures, defined by the
projected density of galaxies. These structures are associated with,
but not centered on, the two brightest Virgo galaxies, M87 and NGC
4472.  We follow Huchra (1985) and refer to these as the M87
sub-cluster and the NGC 4472 sub-cluster.  Table 6 shows that the
mean velocities of the two sub-clusters, measured within a
2\deg~radius, are the same. When higher velocity spirals located at
larger radii from the center are included, the   mean velocity for the
NGC 4472 sub-cluster increases while the mean velocity for the M87
sub-cluster remains approximatively constant. The two sub-clusters
also differ in their ratio of spirals to ellipticals, which is smaller
in the M87 than in the NGC 4472 sub-cluster.

The physical reality of the two sub-clusters is confirmed by X-ray
data.  The Rosat observations of the Virgo cluster (B\"ohringer \etal 1994) show in
beautiful detail the distribution of X-ray emitting gas first detected
by the Einstein and Exosat satellites.  These observations show that
the Virgo cluster core is filled with hot gas, and that the M87 sub-cluster
corresponds to the deepest  potential well.  In contrast to the fact
that neither M87 nor NGC 4472 are at the geometrical center of the
galaxies isopleths, to first order each galaxy is at the center of the
X-ray emitting corona and, consequently, of the dark mass that
dominates the cluster, with M87 having the lion's share. Based on
velocities and Tully-Fisher and fundamental plane distances to 59
early-type and 75 late-type galaxies, Gavazzi \etal  (1999) conclude
that the M87 and NGC 4472 sub-clusters  are at the same distance (a
conclusion also supported by the SBF observations presented in Ajhar
\etal 1999).  

Of the galaxies with Cepheid distances, NGC 4321, NGC 4548 and NGC
4571 are located in the the M87 sub-cluster, while NGC 4535 is
located in the NGC 4472 sub-cluster. The association of NGC 4536 and
NGC 4496A is more problematic, as the galaxies are found over 5\deg~
south of NGC 4472. According to Gavazzi \etal (1999) the two galaxies
are within the NGC 4472 sub-cluster, which they extend farther south
than previous authors, and beyond the region dominated by the X-ray
emission associated with NGC 4472 (even if NGC 4496A is not included
in the Gavazzi \etal survey, its location and its systemic velocity
place it within the sub-cluster). Therefore we include all three
galaxies in the NGC 4472 sub-cluster; in practice this decision is
of little consequence as the three galaxies have very similar Cepheid
distances. 

In addition to the two sub-clusters described above, Gavazzi \etal 
(1999) identify several other clumps, one of which, named the `E
cloud', is located to the east of the M87 sub-cluster and contains
our last Cepheid galaxy, NGC 4639.  Cloud E is found to have the same
recessional velocity as the M87 sub-cluster, but is $\sim 0.4 \pm
0.2$ mag in the background.  In keeping with Huchra's nomenclature, we
rename this region as the  NGC 4649 (M60) sub-cluster from the most
prominent galaxy found here. We would like to stress that while the
physical association of galaxies in the M87 and NGC 4472 sub-clumps
is well established, this is not the case for the NGC 4649
sub-clump, and in fact SBF distances in this region (Ajhar \etal
1999, Ferrarese \etal 2000a) are very heterogeneous and point to a
more complex structure than envisioned by Gavazzi \etal (1999).

We would like to spend a few extra words regarding the confinement  of
NGC 4639 to a different region of the Virgo cluster than M87 and NGC 4472 which,
as described above, are likely to define the cluster's center. 
The issue is particularly important, because the Cepheid
distance modulus to NGC 4639 is $\sim 31.9$ mag, while the remaining
five Cepheid galaxies  have distance modulus $\sim 31.0$ mag, with
very small dispersion. The reasonable doubt here is that  M100, NGC
4494A, NGC 4535, NGC 4536 and NGC 4548 might be at the near side of
the cluster, while NGC 4639 might be at the far side, and therefore
the `true' distance to the Virgo cluster would be better defined by the mean of
all six Cepheid distances, rather than by the mean of the five
`nearby' galaxies. In addition to the (admittedly not very strong)
evidence reported above that NGC 4639 lies in a  clump that seems
to be at slightly larger distances than the region around M87 and NGC
4472, there are two strong reasons to reject NGC 4639 in determining
the distance to the Virgo cluster. The first is based on simple geometrical
arguments: NGC 4639 is located 3\deg~east of M87, which is also the
radius of the  cluster's core as defined by Huchra (1985). If
the cluster is approximatively spherical, then the back to front depth
of the core corresponds to 0.2 mag in distance modulus. NGC 4639 is
$\sim 0.8$ mag more distance than the other Cepheid galaxies, implying
that it is   background to the cluster. The question remains whether
the remaining five Cepheid galaxies are indeed representative of
the Virgo cluster mean distance. The strongest argument in support of this
hypothesis is from B\"ohringer \etal (1997) and Vollmer \etal (1999) 
who found that NGC 4548
shows clear signs of  its interstellar medium being stripped and
distorted as a consequence of the galaxy passing close to the center
of the potential well of the cluster. 

Galaxies with PNLF, TRGB, GCLF and SBF distances are placed in one of
the three  sub-clusters in the Virgo cluster (M87, NGC 4472 and NGC 4649); for those
galaxies which are not included in the Gavazzi study, the
classification was based on redshift and position, and it was
unambiguous in all cases.

Figure 2 shows the Rosat X-ray map of the Virgo cluster (from B\"ohringer \etal
1994).  The straight lines define the regions into which Gavazzi
\etal (1999) divide the cluster. The circles, triangles and squares
show the location of the galaxies with Cepheid, SBF, GCLF and PNLF
distances presented in this paper.
	
\noindent {\bf The Fornax Cluster.}  As for the Virgo cluster, we
follow previous studies of the Fornax cluster region rather than
attempting a new classification. The Fornax cluster is not included in
the N93 catalog.  We consider as certain members of the cluster all
galaxies classified by T88 as belonging to the Fornax cluster +
Eridanus Cloud (cloud number 51), group $-$1. All of these galaxies,
with the exception of NGC 1316, are found in the core of the cluster
within a 2\deg$\times$2\deg~region, and have systemic velocities
between 1350 \kms~and 1950 \kms.  To these galaxies, we added NGC
1373, NGC 1380A and NGC 1380B, which are not included in the T88
catalog, but are found in the same confined region of phase space as
the rest of the galaxies. NGC 1316 is located 3\Deg5 SW of the core,
and classified as a true cluster member by T88. As a consequence, we
also classify NGC 1326A (not part of the T88 catalog) as a true
member, because it lies only half a degree away from NGC 1316, and
shares the same systemic velocity. All of the galaxies discussed above
are also listed as true members of the Fornax cluster by Ferguson
(1989).  ESO358-G6, ESO358-G59, IC 1919, NGC1336, NGC 1339, NGC
1351 and NGC 1419 are found between 2\deg~and 3\deg~ from the
core. T88 discusses only NGC 1339, and places it in group +2, first
level association +1. However, Ferguson (1989) considers all of the
above galaxies as true members, and because we see no reason to
isolate NGC 1339 (which has a systemic velocity of $\sim 1370$ \kms,
very close to the cluster mean), we classify all of the above galaxies
as true members.  Likewise, IC 2006 and NGC 1366 are not discussed by
either T88 or Ferguson (1989), although they are found within 3\deg~of
the cluster core and have systemic velocities very typical for the
cluster.  NGC 1425 and NGC 1344 are 6\deg~and 4.5\deg~north of the
core respectively, and therefore not included in the Ferguson study
(which is limited to the inner 3\Deg5 of the cluster). T88 places NGC
1425 in group $-$0, first level association +1, and NGC 1344 in group
+2, first level association +1.  Following the discussion in Mould \etal
(2000) we list these galaxies as certain members of the Fornax
cluster. All of the galaxies mentioned above are within one Abell
radius of the cluster center (Giovanelli \etal 1997).

\noindent {\bf The Eridanus cluster.}  There are no Cepheid distances
to the Eridanus cluster, however, there are both $I$ and $K'$ SBF
distance to four of its galaxies.  NGC 1395 and NGC 1426 are listed by
T88 as belonging to group 51, cloud -4, and are certain members of the
cluster.  NGC 1407 is also in the core of the cluster, and has a
redshift only 60 \kms~higher than NGC 1395. We consider N1407 a true
member in spite of it being classified in cloud -8 by T88. NGC 1400 is
projected in the cluster core, and it is placed in group +8 by T88. Its
velocity is only 550 \kms, much lower than the cluster mean ($\sim
1600$) \kms, and we therefore consider it as a possible member.

\clearpage

\begin{figure}
\figurenum{1a}
\caption{Finding charts showing all galaxies identified by zcat
(Huchra \& Mader 1998) in the vicinity of each group considered in
this paper, within the velocity range shown in each panel. The coordinates
are at J2000; the galaxy  heliocentric velocity is defined by the
color code as shown in the bar on top of each panel, while the symbol
size is proportional to the galaxy B magnitude reported in Table
1. Galaxies for which a Cepheid distance exists are shown as filled
pentagons; galaxies with SBF, PNLF, TRGB and GCLF distances are shown
as open circles, squares, hexagons and triangles respectively. All
other galaxies are shown as open stars.}
\end{figure}

\begin{figure}
\figurenum{1b}
\caption{As for Figure 1a.}
\end{figure}

\begin{figure}
\figurenum{2}
\caption{The Rosat X-ray map of the Virgo cluster (from B\"ohringer \etal
1994).  The straight lines define the regions into which Gavazzi
\etal (1999) divide the cluster. The circles, triangles and squares
show the location of the galaxies with Cepheid, SBF, GCLF and PNLF
distances presented in this paper.}
\end{figure}

\clearpage



\clearpage

TABLES 2 AND 3 ARE IN SEPARATE FILES, AND NEED TO BE INSERTED HERE

\clearpage

\clearpage

%
                       

\clearpage

\noindent$^{a}${Number of Cepheids used in fitting the PL relation.}

\noindent$^{b}${Cepheid distance modulus. Distance moduli followed by 
a colon are not considered reliable for the reasons described in \S 3. The Cepheid PL relation is calibrated assuming a distance
modulus and reddening to the LMC $\mu=18.50 \pm 0.13$ mag and
$E(B-V)=0.10$ mag respectively. This calibration is common to all
Cepheid distances reported here. Furthermore, all fits to the PL data
are performed by fixing the slope of the relation to the values
determined for the calibrating LMC sample. All Cepheid distances are
scaled adopting a ratio of total to selective absorption $R_V=3.3$ mag.
See Ferrarese et al. (1999) for further details. The error on the distance
moduli are random to the collective sample of Cepheids. Systematic errors
amount to 0.16 mag for the galaxies observed with the $HST$, and 0.13 mag
for the ground based galaxies.}

\noindent$^{c}${References for the Cepheid Distances: 
 1. Tolstoy et al. (1995a); 
 2. Sakai, Madore \& Freedman (1999);
 3. Madore \&  Freedman (1991); 
 4. Freedman, Wilson, \&  Madore (1991); 
 5. This paper, data from Musella, Piotto \&  Capaccioli (1997);
 6. Gallart, Aparicio \& Vilchez (1996);
 7. Sakai, Madore \& Freedman (1996);
 8. Sakai, Madore \&  Freedman (1997); 
 9. This paper, with data from Welch et al. (1987); 
 10. Silbermann et al. (1996); 
 11. Prosser et al. (1999); 
 12. Silbermann et al. (1999); 
 13. Mould et al. (1999); 
 14. Tolstoy et al. (1995b);
 15. Freedman et al. (1994); 
 16. Hoessel, Saha \& Danielson (1998).}
 17. Kelson (1999); 
 18. Sakai et al. (1999); 
 19. Graham et al. (1997); 
 20. Gibson et al.  (1999), for NGC 4496A, NGC 5253, NGC 4536, IC 4182 and NGC 4639, the data is also described in Saha et al. (1994, 1995, 1996a, 1996b, 1997, 1999);
 21. Turner et al. (1998); 
 22. Gibson et al. (1998); 
 23. Ferrarese  et al. (1996); 
 24. Graham et al. (1998); 
 25. Pierce et al. (1994); 
 26. Macri et al. (1999); 
 27. Kelson (1996);  
 28. Hughes (1998); 
 28. Freedman et al. (1992); 
 30. Phelps et al. (1998); 
 31. Ferrarese et al. (1998); 
 32. Rawson et al. (1997);
  
\noindent$^{d}${I magnitude for the tip of the RGB. The magnitudes are not corrected for either internal or
Galactic extinction.}

\noindent$^{e}${References for the TRGB Distances:
 1. Armandroff et al. (1998);
 2. Hopp et al. (1999);
 3. Grebel \& Guhathakurta (1998);
 4. Lee \& Byun (1999);
 5. Sakai, Madore \& Freedman (1998);
 6. Freedman (1988); 
 7. Caputo et al. (1999);
 8. Tolstoy et al. (1998);
 9. Lee (1995a); 
10. Lee (1995b); 
11. Reid, Mould \& Thompson (1987); 
12. Mould, Kristian \& da Costa (1984); 
13. Mould \& Kristian (1986); 
14. Lee (1993); 
15. Minniti et al. (1999);
16. Gallart, Aparicio \& Vilchez (1996);
17. Sakai, Madore \& Freedman (1996); 
18. Sakai, Madore \& Freedman (1997); 
19. Lee, Freedman, \& Madore (1993); 
20. Caldwell et al. (1998);
21. Sakai \& Madore (1999);
22. Sakai et al. (1997); 
23. Harris et al. (1998); 
24. Ferguson, Tanvir \& Hippel (1998);
25. Soria et al. (1996); 
26. Elson (1997).}

\noindent$^{f}${Number of Planetary Nebulae used in fitting the
PNLF. This excludes PNe that are fainter than the completeness limit
adopted for the data.}

\noindent$^{g}${Apparent cutoff $\lambda$5007 magnitude for the
PNLF. Magnitudes are not corrected for extinction, either internal or
Galactic. Magnitudes followed by a colon are not considered reliable for the reasons described in \S 3.}

\noindent$^{h}${References for the PNLF $m^*_{5007}$ cutoff magnitudes:
 	1. Jacoby, Walker \&  Ciardullo (1990); 
 	2. Ciardullo et al. (1989); 
 	3. Richer \& McCall (1992); 
 	4. Ciardullo, Jacoby \&  Harris (1991); 
 	5. McMillan, Ciardullo \&  Jacoby (1993); 
 	6. Jacoby et al. (1989); 
 	7. Feldmeier, Ciardullo \&  Jacoby (1997); 
	8. Ciardullo, Jacoby \&  Ford (1989); 
 	9. Jacoby, Ciardullo \&  Harris (1996); 
 	10. Jacoby, Ciardullo \&  Ford (1990); 
 	11. Ciardullo et al. (1998);
 	12. McMillan, Ciardullo \&  Jacoby (1994); 
 	13. Hui et al. (1993); 
 	14. Phillips et al. (1992); 
	15. Feldmeier, Ciardullo \&  Jacoby (1996);
 	16. Soffner et al. (1996);
 	17. Ciardullo, Jacoby\&  Tonry (1993); 
 	18. Ford et al. (1996).}

\noindent$^{i}${The apparent $V-$band magnitude of the turnover of the
globular cluster luminosity function, uncorrected for reddening.}

\noindent$^{j}${References for the GCLF Distances:
1. Reed, Harris \& Harris (1994);
2. Blakeslee \&  Tonry (1996); 
3. Elson et al. (1998); 
4. Della Valle et al. (1998);
5. Ostrov, Forte \& Geisler (1998);
6. Grillmair et al. (1999);
7. Richtler et al. (1992); 
8. This paper, data from Perelmuter \& Racine (1995); 
9. Pritchet \&  van den Bergh (1985); 
10. Forbes (1996a); 
11. Fleming et al. (1995); 
12. Neilsen, Tsvetanov \&  Ford (1997); 
13. Kundu et al. (1999);
14. Harris et al. (1991); 
15. Lee, Kim \& Geisler (1998); 
16. Baum et al. (1996); 
17. Kundu \& Whitmore (1998);
18. Durrell et al. (1996); 
19. Forbes (1996b); 
20. Fischer et al. (1990);
21. Madejsky \& Rabolli (1995);
22. Forbes, Brodie \&  Huchra (1996).}

\noindent$^{k}${SBF magnitudes. All values are not corrected for
extinction. The $I$-SBF fluctuation magnitudes reported here must be 
cited as Ajhar et al. (1999).
}

\noindent$^{l}${References for the SBF Distances:
1. Ajhar et al. (1999);
2. Ajhar et al. (1997);
3. Jensen, Tonry \& Luppino (1998);
4. Pahre et al. (1999);
5. Pahre \& Mould (1994);
6. Neilsen, Tsvetanov \&  Ford (1997), not used for distance determinations since the galaxy's mean color is not given.;
7. Thomsen et al. (1997);
8. Lauer et al. (1998).}

\noindent$^{m}${Adopted zero point for the Cepheid PL relations.}

\noindent$^{n}${The distance modulus is determined as an upper limit
from the magnitude of the brightest PN in a  sample of only seven
objects.}

\noindent$^{o}${The published value is $T_1=23.0\pm0.15$. The $C-T_1$
color is also available, allowing to  convert the $T_1$ magnitude to
$V$ using the formulas by Geisler (1996)}

\noindent$^{p}${The published value is $T_1=R=23.3\pm0.1$. The $C-T_1$
color is also available, allowing to  convert the $T_1$ magnitude to
$V$ using the formulas by Geisler (1996)}

\noindent$^{q}${For the distance to M101, we adopt the value derived
by Kelson et al. (1996). Stetson et al. (1998)
also derived a distance to M101 based on HST/WFPC2 observations of a
different field of the galaxy.}

\noindent$^{r}${Distance modulus determined using the cumulative PNLF.}

\noindent$^{s}${The quoted values is calculated from the published
unreddened values using $A_I=0.05$.}

\noindent$^{t}${These distances have been modified from the published value to account
for period incompleteness bias (see \S 3 and Table 4  for further details).}

\noindent$^{u}${Distances from Gibson's et al. (1999), on the Hill et al. (1998) zero point (from column 2 of Table 4 of Gibson et al. 1999).}